\documentclass[twocol]{ametsocV6.1}
\usepackage{subcaption} 
\usepackage{color,soul}

\title{The upwelling source depth distribution and its response to wind stress and stratification}

\authors{Elle Weeks\aff{a}\correspondingauthor{Elle Weeks, elleweeks@g.harvard.edu},
Martin Losch\aff{b} \&
Eli Tziperman\aff{a,c} 
}

\affiliation{\aff{a}{School of Engineering and Applied Sciences, Harvard University, Cambridge, Massachusetts}\\
\aff{b}{Alfred-Wegener-Institut f\"ur Polar-und Meeresforschung, Bremerhaven, Germany}\\
\aff{c}{Department of Earth and Planetary Sciences, Harvard University, Cambridge, Massachusetts}
}

\abstract{Coastal upwelling, driven by alongshore winds and characterized by cold sea surface temperatures and high upper-ocean nutrient content, is an important physical process sustaining some of the oceans' most productive ecosystems. To fully understand the ocean properties in eastern boundary upwelling systems, it is important to consider the depth of the source waters being upwelled, as it affects both the SST and the transport of nutrients toward the surface. Here, we construct an upwelling source depth distribution for parcels at the surface in the upwelling zone. We do so using passive tracers forced at the domain boundary for every model depth level to quantify their contributions to the upwelled waters. We test the dependence of this distribution on the strength of the wind stress and stratification using high-resolution regional ocean simulations of an idealized coastal upwelling system. We also present an efficient method for estimating the mean upwelling source depth. Furthermore, we show that the standard deviation of the upwelling source depth distribution increases with increasing wind stress and decreases with increasing stratification. These results can be applied to better understand and predict how coastal upwelling sites and their surface properties have and will change in past and future climates.}

\begin{document}

\maketitle

%
%
%
%
%
%

%

\section{Introduction}
\label{sec:introduction}

Wind-driven coastal upwelling plays an important role in determining the dynamics and physical characteristics of coastal upwelling systems, affecting both the sea surface temperature (SST) and biological productivity. Due to the high concentration of nutrients transported by the upwelling toward the surface, these coastal upwelling sites are some of the world's most productive ecosystems, showing both high primary productivity and ecological biodiversity \citep{Hutchings-Pitcher-Probyn-et-al-1995:chemical, Chavez-Messie-2009:comparison}. Despite their biological importance, eastern boundary coastal upwelling systems remain poorly modeled by global climate models \citep{Richter-2015:climate}, and gaps remain in our understanding of what sets the surface ocean properties such as SST and nutrient content. Specifically, to fully understand what sets the surface properties of the ocean in and around coastal upwelling sites, it is necessary to consider what determines the source depth of the upwelled water \citep{He-Mahadevan-2021:how}. In fact, one expects the upwelled water to arrive from more than one single depth, and our interest here is in developing an approach to efficiently estimate both the mean upwelling source depth and the full distribution of sources from which the upwelled water originates.

In coastal upwelling sites, alongshore winds drive offshore Ekman transport and, by continuity, the rise of water from depth along the coastline. Strong upward vertical flow occurs in a narrow region near the coast on the same length scale as the Rossby deformation radius. The strong offshore Ekman flow is balanced by a weaker and broader onshore return flow below the surface Ekman layer. An alongshore coastal jet flows equatorward, in the same direction as the wind, along the coastline near the surface. At depth, a poleward undercurrent develops, flowing in the opposite direction as the equatorward surface jet. These features of coastal upwelling sites are regularly seen in observations \citep{Huyer-1983:coastal, Spall-Schneider-2016:coupled, Zaba-Rudnick-Cornuelle-et-al-2020:volume} and realistic simulations \citep{Capet-McWilliams-Molemaker-et-al-2008:mesoscale}. These features were explained using multiple modeling approaches, including both idealized linear \citep{Pedlosky-1974:longshore, McCreary-1981:linear} and nonlinear \citep{Pedlosky-1978:nonlinear, Choboter-Samelson-Allen-2005:new} as well as steady \citep{Pedlosky-1978:inertial} and time-dependent \citep{Samelson-2017:time}.

Furthermore, as isopycnals outcrop near the coast as a result of coastal upwelling, strong fronts are generated, and the resulting baroclinic instabilities in the mixed layer lead to submesoscale turbulence. These submesoscale eddies then drive a circulation that acts to restratify the upper ocean, countering the effects of the wind-driven Ekman circulation \citep{marshall2003residual, Capet-Marchesiello-McWilliams-2004:upwelling, Thomsen-Capet-Echevin-2021:competition}. The strength of the eddy-driven restratification has been shown to scale with the horizontal density gradient, mixed layer depth squared, and the inertial period \citep{fox2008parameterization}. The width of the upwelling zone, where deep waters are advected upwards and isopycnals outcrop, is expected to be proportional to the baroclinic Rossby radius of deformation \citep{Pedlosky-1978:inertial, Lentz-Chapman-2004:importance, He-Mahadevan-2021:how}. \cite{Lentz-Chapman-2004:importance} showed that the slope of isopycnals in the upwelling zone was $0.25f/N$ implying a length scale of $L=4{ND_s}/{f}$ for the width of the upwelling front, where $f$ is the Coriolis frequency, $N$ is the stratification, and $D_s$ is the depth of the mixed layer. Independently, \cite{Spall-Schneider-2016:coupled} showed that the decay length scale for the SST anomaly is ${\tau c_p}/{\Gamma f}$, where $\tau$ is the wind stress, $c_p$ is the specific heat of water, and $\Gamma$ is the atmospheric-ocean heat exchange sensitivity to temperature difference in W/m$^2$K.

An extensive body of previous work has explored the dynamics of coastal upwelling sites and how the SST relates to the strength of the surface wind stress and the strength of the stratification in the ocean \citep{Capet-Marchesiello-McWilliams-2004:upwelling, Chavez-Messie-2009:comparison, Spall-Schneider-2016:coupled, Miller-Tziperman-2017:effect, Zaba-Rudnick-Cornuelle-et-al-2020:volume}. Yet, in spite of substantial work on upwelling dynamics, it is still not clear what controls the distribution of source depths of the upwelled waters. Some existing studies calculate the mean upwelling source depth and suggest that this source depth may depend on the magnitude and spatial structure of the wind stress, the buoyancy gradient (stratification), the Coriolis frequency, and the bottom topography \citep{Lentz-Chapman-2004:importance, Song-Miller-Cornuelle-et-al-2011:changes, Jacox-Edwards-2011:effects, Jacox-Edwards-2012:upwelling, He-Mahadevan-2021:how}. \cite{Jacox-Edwards-2011:effects} investigated how the shelf slope and stratification affect the source depth in a two-dimensional model. They estimated the source depth by introducing a single passive tracer with initial conditions that increase linearly with depth to diagnose the source depth, and found that a steep shelf slope and weak stratification resulted in the greatest source depth. \citet[][hereafter referred to as HM]{He-Mahadevan-2021:how} derived a scaling relationship for the mean upwelling source depth in terms of the wind stress, stratification, and Coriolis frequency by considering a balance between the wind-driven overturning circulation and the eddy-driven restratification. They validated the theorized relationship using three-dimensional numerical simulations with periodic boundary conditions, ignoring the effects of bottom topography or a surface heat flux, and estimated the mean source depth using passive tracers initialized separately for every model depth level. Notably, the scaling relationship described by HM for the mean source depth had the same functional dependence on wind stress, stratification, and Coriolis frequency as a scaling for the depth of the wind-driven mixed layer in the open ocean suggested by the simple model described by \cite{Pollard-Rhines-Thompson-1973:deepening}. 

Understanding the mean source depth and the source depth distribution requires a different modeling approach than used in many prior studies; it is important that the modeling strategy allows the source depth to be determined by the dynamics rather than be prescribed. Previous studies have typically used one of three approaches for modeling coastal upwelling sites: (1) a 2-dimensional modeling domain \citep{Lentz-Chapman-2004:importance, Jacox-Edwards-2011:effects, Jacox-Edwards-2012:upwelling}, (2) periodic boundary conditions in the alongshore direction \citep{He-Mahadevan-2021:how, Thomsen-Capet-Echevin-2021:competition}, or (3) a realistic model configuration requiring boundary conditions prescribed from observations \citep{Capet-Marchesiello-McWilliams-2004:upwelling, Song-Miller-Cornuelle-et-al-2011:changes}. None of the previous modeling studies of coastal upwelling sites have allowed for a statistical steady state to be reached while also allowing the source depth to evolve freely. The 2-D modeling studies are limited by not allowing for any alongshore variability. Periodic boundary conditions in the alongshore direction allow for some alongshore variability and for the natural development of eddies but, along with 2-D models, introduce unique problems for modeling coastal upwelling systems. The offshore Ekman transport is balanced by an onshore return flow in the ocean interior that is generally considered to be a geostrophic current driven by an alongshore pressure gradient \citep{Huyer-1983:coastal}. However, in both 2-D models and models with periodic boundary conditions, no alongshore gradients can develop. One solution is to prescribe an alongshore pressure gradient force within a certain depth range in the interior, yet this solution prescribes and directly controls the depth of the return flow and may have consequences for the upwelling source depth \citep{Thomsen-Capet-Echevin-2021:competition}. Due to these constraints, the simulations used by \cite{Jacox-Edwards-2011:effects, Jacox-Edwards-2012:upwelling} and \cite{He-Mahadevan-2021:how} were not run to a statistical steady state. Realistic modeling studies may be run to a steady state using prescribed boundary conditions for temperature, salinity, and inflow/outflow derived from ocean reanalysis products; however, these prescribed boundary conditions do not allow the source depth to adjust freely.

To estimate the mean source depth in a given modeling regime, numerous methods have been used. Previous approaches include identifying the depth at which the density of surface water parcels along the coast matches the initial/offshore vertical density profile \citep{carr2003production}, identifying the depth of the strongest return flow \citep{Davis-2010:coastal}, using passive tracers to track the initial depth of water parcels \citep{Chhak-Di-2007:decadal, Song-Miller-Cornuelle-et-al-2011:changes,He-Mahadevan-2021:how}, or using Lagrangian analyses to track the origin of water parcels \citep{mason2012lagrangian,ragoasha2019lagrangian}. Additionally, beyond the calculation of the mean upwelling source depth, it is valuable to be able to calculate the full distribution of depths from which water arrives at the surface of the upwelling zone and to predict how it might change in a different climate, because it affects both the resulting coastal SST and the distribution of nutrients in the upper ocean. However, to our knowledge, no previous work addressed this issue and characterized the full distribution of the upwelling source depth. 

In this work, we address three issues related to the source depth of upwelling zones. First, we present an idealized numerical modeling approach that enables the evolution of the source depth to be freely determined by the model. Our model includes the effects of non-flat bottom topography and a surface heat flux, and simulations in this work are run to a statistical steady state. Second, we introduce a single passive depth tracer that can be used to accurately and efficiently estimate the mean upwelling source depth, including its spatial and temporal variability. We use this tracer to investigate the response of the mean source depth to spatially and temporally uniform wind stress and linear stratification. We compare the results for the mean upwelling source depth to those of previous studies where more restrictive modeling assumptions were made \citep{He-Mahadevan-2021:how}. Finally, we introduce the idea that upwelling arrives from multiple depths, and to be fully described requires estimating the distribution of depth sources that feed the surface water of the upwelling zone. We present an approach to estimate this distribution using a set of passive tracers forced separately at the boundary for every depth level (somewhat similar to HM, except that they used a set of initialized tracers to calculate the mean depth source rather than the distribution of depth sources). We characterize the full upwelling source depth distribution and its response to the strength of the wind stress and stratification. We quantify the effect of wind stress and stratification on the width of the upwelling source depth distribution as measured by its standard deviation. To our knowledge, this is the first work to consider the upwelling source depth distribution rather than only the mean source depth.

In the following sections, we begin by describing a numerical model for an idealized coastal upwelling region and introducing methods for quantifying the upwelling source depth distribution (Section~\ref{sec:methods}). In Section~\ref{sec:results}, we first demonstrate that our model is able to recreate known upwelling dynamics while allowing the upwelling source depth to evolve freely (Section~\ref{sec:results}\ref{sec:results:novel-simulations-aspects}), then show support for previous results for the scaling of the mean upwelling source depth using a single depth tracer (Section~\ref{sec:results}\ref{sec:results:mean-depth}) and, finally, discuss new results characterizing the full upwelling source depth distribution (Section~\ref{sec:results}\ref{sec:results:distribution}). We discuss and conclude in Section~\ref{sec:conclusion}.

\section{Methods}
\label{sec:methods}

\subsection{Numerical model}
\label{sec:methods:model}

We perform high-resolution, regional ocean simulations of an idealized coastal upwelling system using the MIT general circulation model \citep[MITgcm,][]{Marshall-Adcroft-Hill-et-al-1997:finite, Marshall-Hill-Perelman-et-al-1997:hydrostatic, Adcroft-Hill-Campin-et-al-2004:overview, alistair2018mitgcm}. We model the upwelling system in a rectangular domain on a $\beta$-plane centered at 37\textdegree N (approximately the midlatitude of the California Current System). The computational domain is 600 km (cross-shore, $x$) by 1200 km (along-shore, $y$) with a maximum depth of 1000 m. The horizontal resolution is 2 km, and there are 50 vertical levels ranging in depth from 2.5 m at the surface to 75 m at the bottom. An idealized bathymetry that is uniform in the alongshore direction with the coastline on the eastern boundary is motivated by the California continental slope and has the following functional form:
\begin{equation}
    h(x)=h_{max} \left(1+\tanh \left( \frac{x-x_s}{L_s} \right) \right),
\end{equation}
with $L_s = 20$ km, $x_s = 35$ km, $h_{max} = 20$ m \citep{Thomsen-Capet-Echevin-2021:competition}. No-slip boundary conditions are enforced along the sides and bottom of the domain. Salinity is not simulated, and the density varies with temperature according to a linear equation of state (thermal expansion coefficient estimated at a temperature of 20\textdegree C, $\alpha=2 \times 10^{-4}$ 1/K). In the horizontal, mixing is set by a constant biharmonic eddy viscosity and diffusivity, which are fixed at $2.5 \times 10^7$ and $1 \times 10^6$ $m^4/s$, respectively in all simulations. Subgridscale vertical mixing is represented with the K-Profile Parameterization \citep{large1994oceanic}. Outside the KPP boundary layer, the vertical eddy viscosity is set to $1 \times 10^{-4}$ $m^2/s$ and the vertical eddy diffusivity is set to $1 \times 10^{-5}$ $m^2/s$.

The model is initialized with a horizontally uniform and vertically stratified temperature profile and started from a state of rest. The model is forced with temporally and spatially uniform wind stress over the meridional middle third of the domain. The wind stress decays meridionally to zero away from the center third, following a hyperbolic tangent over the northern and southern thirds of the domain. This allows the upwelling dynamics to develop in the middle third of the domain with ample buffer to the northern and southern boundaries.

At the surface, a prescribed heat flux and weak SST restoring are applied such that the climatological model mean SST remains relatively constant while also being allowed to develop zonal temperature gradients. The surface heat flux takes the following form:
\begin{equation}
    H_{surface}(x,y,t)=\gamma(T^*(y)-T(x,y,t))+H_0(y),
\end{equation}
where $\gamma$= 1/7.5 days, $T^\ast(y)$ is a restoring SST profile, $T$ is the model SST, and $H_0(y)$ is the prescribed heat flux. The restoring climatological SST, $T^\ast(y)$, has a constant meridional temperature gradient equal to that of the average SST gradient of the California State Estimate Short-term State Estimation (CASE-STSE) reanalysis product temperature over the domain spanned diagonally from 235\textdegree E, 34.5\textdegree N to 232\textdegree E, 39\textdegree N. The time-independent part of the heat flux, $H_0(y)$, was computed from a simulation forced with strong SST restoring ($\gamma_{strong} = 1/2.5$ days). The model restoring term, $\gamma_{strong}(T^\ast(y)-T)$, was then temporally and zonally averaged over the western third of the domain and converted to the zonally constant heat flux, $H_0(y)$.

The computational domain is closed to the north, south, and west; at the western boundary, there is a 50 km wide sponge layer (25 grid points). We define the physical domain as everything east of the western boundary sponge layer. In this way, our boundary conditions allow flow into/out of the physical domain via vertical and horizontal flows within the sponge layer. We perform most of the analysis in the meridional middle third of the physical domain, where inflow and outflow from the north and south may occur freely as required by the dynamics. This allows us to analyze the upwelling dynamics in a region where they are able to develop without being influenced by the closed boundaries of the computational domain. The meridional average value of the temperature in the sponge layer along the boundary at a given depth, $z$, is restored to $T_0(z)$. This restoring replaces the more common point-wise restoring used in previous studies and allows meridional temperature gradients to freely develop. This restoring term takes the following form:
\begin{equation}
    H_{wb}(x_{sponge},z,t) = \gamma_{sponge} \left(T_0(z)-\frac{1}{L_y}\int_0^{L_y}T(x_{sponge},y,t)dy\right),
\end{equation}
where, $L_y$ is the alongshore length of the domain, $x_{sponge}$ is the longitudinal location within the sponge layer, and $\gamma_{sponge}$ ranges from 1/2.5 days along the boundary to 1/125 days at the inner edge of the sponge layer bordering with the physical domain.

We run nine primary experiments varying the strength of the constant linear stratification and uniform wind stress. We test the same parameter values for stratification ($N$) and alongshore wind stress ($\tau_y$) used by HM: $N^2 = 1\times 10^{-5}$, $5.5\times 10^{-5}$, $1\times 10^{-4}$ 1/s$^2$ and $\tau_y = 1\times10^{-2}$, $5.5 \times10^{-2}$, $1\times 10^{-1}$ N/m$^2$. We run six additional experiments with constant wind stress and stratification to obtain better coverage of the parameter space over a range of source depths. (See Table~\ref{table:sims} for a description of all simulations.) The model is run for nine years, and temporal averages of temperatures and velocities are taken over the final five years after the model has reached a statistical steady state. Passive tracers (described below) are introduced after nine years, allowed to spin up for one year, and temporal averages of tracers are taken over two years after spin up. 

\begin{table}[!tbh] 
\centering
 \begin{tabular}{||l | c c||} 
 \hline
 \bf{Experiment} & \bf{Wind stress} & \bf{Stratification} \\ 
   & $\tau$ (N/m$^2$) & N$^2$ (s$^{-2}$)\\ [0.5ex] 
 \hline\hline
 low wind, low strat & $1\times10^{-2}$ & $1\times 10^{-5}$ \\
 low wind, med strat & $1\times10^{-2}$ & $5.5\times 10^{-5}$ \\
 low wind, high strat & $1\times10^{-2}$ & $1\times 10^{-4}$ \\
 med wind, low strat & $5.5 \times10^{-2}$ & $1\times 10^{-5}$ \\
 med wind, medium strat & $5.5 \times10^{-2}$ & $5.5\times 10^{-5}$ \\
 med wind, high strat & $5.5 \times10^{-2}$ & $1\times 10^{-4}$ \\
 high wind, low strat & $1\times 10^{-1}$ & $1\times 10^{-5}$ \\
 high wind, med strat & $1\times 10^{-1}$ & $5.5\times 10^{-5}$ \\
 high wind, high strat & $1\times 10^{-1}$ & $1\times 10^{-4}$ \\ [0.5ex] 
 \hline
 med-low wind, med strat & $2.5\times 10^{-2}$ & $5.5\times 10^{-5}$ \\
 med-high wind, med strat & $7.5 \times10^{-2}$ & $5.5\times 10^{-5}$ \\
 med wind, med-low strat & $5.5 \times10^{-2}$ & $2.5 \times10^{-5}$ \\
 med wind, med-high strat & $5.5 \times10^{-2}$ & $7.5 \times10^{-5}$ \\
 med-high wind, med-low strat & $7.5 \times10^{-2}$ & $2.5 \times10^{-5}$ \\
 med-low wind, med-high strat & $2.5\times 10^{-2}$ & $7.5 \times10^{-5}$ \\ [1ex] 
 \hline
 \end{tabular}
  \caption{Summary of the values used for the alongshore wind stress and linear stratification in the numerical simulations. Nine primary experiments listed first with the six additional experiments below.}
  \label{table:sims}
\end{table}

\begin{figure*}[!tb]
     \centering
     \begin{subfigure}[b]{0.31\textwidth}
         \centering
         \includegraphics[width=\textwidth]{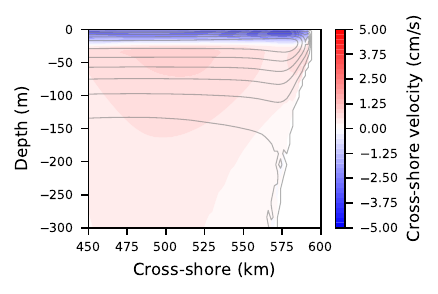}
         \caption{}
         \label{fig:u}
     \end{subfigure}
     \hspace{0.01\textwidth}
     \begin{subfigure}[b]{0.31\textwidth}
         \centering
         \includegraphics[width=\textwidth]{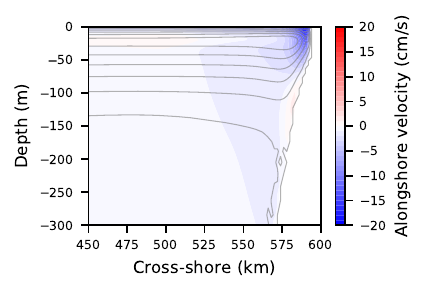}
         \caption{}
         \label{fig:v}
     \end{subfigure}
     \hspace{0.01\textwidth}
     \begin{subfigure}[b]{0.31\textwidth}
         \centering
         \includegraphics[width=\textwidth]{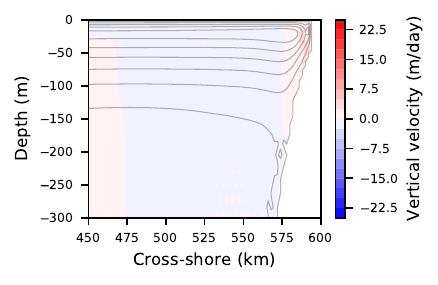}
         \caption{}
         \label{fig:w}
     \end{subfigure}
        \caption{Temporal-mean, alongshore-averaged (a) cross-shore velocity showing strong offshore Ekman transport near the surface and a weaker onshore return flow in the interior, (b) alongshore velocity showing a strong equatorward surface jet and poleward undercurrent, and (c) vertical velocity showing strong upward transport in the upwelling zone for experiment with $N^2=5.5\times 10^{-5}$ and $\tau=5.5\times 10^{-2} $. The Eulerian mean stream function, defined for the alongshore average circulation over the full domain as $\psi=\int_{-1000}^0 \bar{u}^y dz$, is shown in grey contours in (a), (b), and (c).}
        \label{fig:vel}
\end{figure*}

\begin{figure*}[tbh]
     \centering
     \begin{subfigure}[b]{0.4\textwidth}
         \centering
         \includegraphics[width=\textwidth]{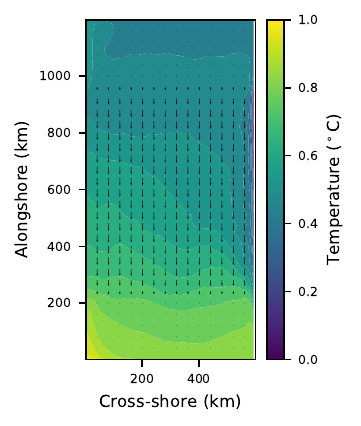}
         \caption{}
         \label{fig:sst}
     \end{subfigure}
     \hspace{0.01\textwidth}
     \begin{subfigure}[b]{0.4\textwidth}
         \centering
         \includegraphics[width=\textwidth]{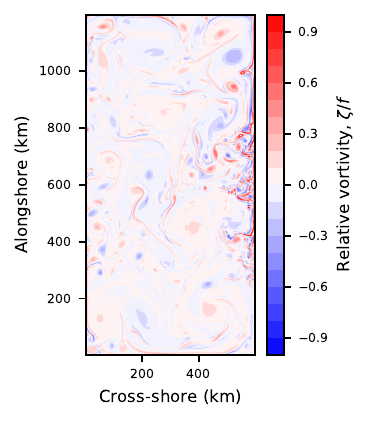}
         \caption{}
         \label{fig:vort}
     \end{subfigure}
    \caption{Surface fields for experiment with $N^2=5.5\times 10^{-5}$ and $\tau=5.5\times 10^{-2}.$ (a) Temporal-mean sea surface temperature showing a strong cooling the upwelling zone. Overlain arrows indicate direction and strength of the applied wind stress. (b) Instantaneous relative vorticity scaled by the Coriolis frequency $(\zeta/f)$. Maximum values of $(\zeta/f)$ close to one indicate a strong submesoscale eddy field.}
    \label{fig:surf}
\end{figure*}

Our idealized simulations recreate the mean state and circulation patterns of the known upwelling dynamics. The climatological mean circulation and SST patterns for one experiment with medium wind stress and stratification are shown in Figs.~\ref{fig:vel} and \ref{fig:surf}. A strong offshore transport near the surface in the Ekman layer is seen in Fig.~\ref{fig:u} as a negative cross-shore velocity in the upper 25 m of the ocean. The magnitude of the offshore transport is consistent with the expected ${\tau}/{\rho f}$. There is a weaker-velocity onshore return flow in the interior that compensates for the offshore surface flow (Fig.~\ref{fig:u}). In the alongshore direction, there is the expected equatorward surface jet near the coast that weakens offshore and a weak poleward undercurrent confined close to the coastline (Fig.~\ref{fig:v}). Strong upward vertical velocities are generated in the upwelling zone close to the coast, reaching amplitudes of up to $2\times 10^{-2}$ cm/s or 25 m/day (Fig.~\ref{fig:w}). These circulation features are consistent with observations and previous realistic modeling studies \citep{Capet-Marchesiello-McWilliams-2004:upwelling, Capet-McWilliams-Molemaker-et-al-2008:mesoscale, Davis-2010:coastal, Zaba-Rudnick-Cornuelle-et-al-2018:annual, Zaba-Rudnick-Cornuelle-et-al-2020:volume}.

The sea surface temperature exhibits a characteristic cooling along the eastern coastal boundary in the meridional middle third of the domain where the upwelling-favorable wind forcing is applied (Fig.~\ref{fig:sst}). The strongest cooling signal can be seen in a narrow band close to the coast where coastal upwelling is active; the cooling signal also propagates westward and can be seen up to hundreds of kilometers offshore \citep{Spall-Schneider-2016:coupled}. The cold anomaly observed in the upwelling zone relative to the western boundary of the domain ranges from 0.5\textdegree{}C to 3\textdegree{}C depending on the strength of the wind stress and the stratification in the simulation. Fig.~\ref{fig:sst} shows the SST pattern for one experiment with medium wind stress and stratification where the temperature across the domain cools by 1.5\textdegree{}C on average from west to east. The model also develops a strong submesoscale eddy field with a Rossby number ($\zeta/f$) of up to and even slightly larger than one (Fig.~\ref{fig:vort}). The relative vorticity is strongest in the upwelling zone along the eastern coastal boundary where the vertical outcropping of isopycnals occurs and where we expect submesoscale eddies to be most active.

\subsection{The mean source depth tracer}
\label{sec:methods:mean}

We review previously used methods for estimating the mean source depth in the introduction (Section~\ref{sec:introduction}). Here, we introduce a novel way of estimating the mean source depth in a way that accounts for both mixing and advection that is also computationally efficient. We estimate the mean source depth using a single passive tracer defined as follows. After the model is run to steady state, the value of the mean-depth tracer is initialized in each model depth level to be the mean depth of that level. Explicitly, for each of 50 vertical levels in the model $k=1,...,50$, the mean depth tracer $C_d$ is initialized as follows:
\begin{equation}
    C_d(x,y,k) = d(k),
\end{equation}
where $d(k)$ is the depth of the center of vertical level $k$. The mean-depth tracer is then forced at the western boundary during the model run with a restoring term that takes the following form for $x<50$ m (grid cells in the western boundary sponge layer):
\begin{equation}
    \frac{\partial C_d(x,y,k)}{\partial t} = \frac{1}{\delta t} \left(d(k)-C_d(x,y,k)\right),
\end{equation}
where $\delta t$ is the model timestep. After spin-up, the value of this depth tracer in the upwelling zone represents the mean upwelling source depth accounting for both advection and mixing. We choose to characterize the source depth in the upwelling zone by analyzing the set of grid cells that are in the uppermost vertical level (2.5 m) and are most proximal to the coast (within 2 km). We compute the mean upwelling source depth as the average value of the mean-depth tracer in this set of grid cells. Computing averaged quantities over grid cells in a wider or deeper upwelling zone was also tested but did not significantly alter the results. More generally, the value of the mean-depth tracer provides an estimate of the depth in the source region (the western boundary sponge layer in this case) from which water parcels originate. This means that the mean-depth tracer can similarly be used to define the mean source depth for any given fluid parcel, not just those within the upwelling zone.

\subsection{Estimating the source depth distribution}
\label{sec:methods:distribution}

We investigate the full source depth distribution using a unique passive tracer for each model depth level to track their contribution to the make-up of fluid parcels. Similarly to HM, we introduce 50 passive tracers, one for each vertical level in the model. However, where HM only initialized these tracers and computed the mean upwelling source depth, we force these tracers at the boundary (the source region) and construct a full source depth distribution. After the model is run to a steady state, the tracer concentrations in each grid cell are initialized to have a concentration of 1 for the tracer corresponding to its initial depth and 0 otherwise.
Explicitly, for each of 50 vertical levels in the model $k=1,...,50$, a unique depth tracer, $C_k$, is initialized as follows:
\begin{equation}
C_k(x,y,k',t_0) = 
\begin{cases}
        1 & \text{if } k'=k, \\
        0 & \text{if } k'\neq k. 
    \end{cases}
\end{equation} 
The tracer restoring during the model run takes the following form for $x<50$m (grid cells within the western boundary sponge layer):
\begin{equation}
\frac{\partial C_k(x,y,k)}{\partial t} = 
\begin{cases}
        \frac{1}{\delta t}(1-C_k(x,y,k)) & \text{if } k'=k, \\
        \frac{1}{\delta t}(0-C_k(x,y,k)) & \text{if } k'\neq k. 
    \end{cases}
\end{equation}  

The resulting concentration of each tracer in the upwelling zone provides a distribution of the source depths from which the upwelled water originates. Any upwelling zone grid box contains different concentrations of the tracers that originate at different depths in the source region within the sponge layer. From this distribution of tracers, we can compute both the mean and the distribution, including the standard deviation, of the upwelling source depth. The distribution of source depths is simply given by the concentrations: if the concentration of a tracer initialized at a level $k$ is $C_k$, then the fraction, $C_k$, of the water in this grid box comes from that level. 

The mean source depth for a given fluid parcel, which was calculated above using the mean-depth tracer can equivalently be calculated using the 50 passive tracers as follows:
\begin{equation} \label{eq:Ds_mean}
    D_s=\frac{\sum_{k=1}^MC_kd(k)}{\sum_{k=1}^MC_k},
\end{equation}
where $M$ = 50 is the number of tracers/vertical levels in the model, $d(k)$ is the depth of the center of vertical level $k$, and $C_k$ is the concentration of tracer $k$ in the water parcel \citep{He-Mahadevan-2021:how}. The standard deviation of the source depth distribution, which we use to quantify the width of the distribution, is calculated as follows:
\begin{equation} \label{eq:Ds_std}
    \sigma(D_s) = \sqrt{\frac{\sum_{k=1}^MC_kd(k)^2}{\sum_{k=1}^MC_k}-D_s^2}.
\end{equation}
Similarly to the use of the mean-depth tracer to compute the source depth for any water parcel in the model domain, we may also use the set of 50 passive tracers to construct a source depth distribution and estimate its mean and standard deviation for any water parcel. This distribution represents the different depths within the source region from which that fluid parcel originated from.

\section{Results} 
\label{sec:results}

First, having validated our idealized numerical model by demonstrating that it recreates the known characteristics and circulation patterns of a coastal upwelling site, we illustrate how our novel boundary condition formulation allows the source depth to evolve freely while in a statistical steady state (Section~\ref{sec:results}\ref{sec:results:novel-simulations-aspects}). We then present results for the mean upwelling source depth using the mean-depth tracer in Section~\ref{sec:results}\ref{sec:results:mean-depth}. We consider the western boundary of our regional model to be the source of the upwelled water and are interested in quantifying the source depth of the water that eventually upwells to the surface at the eastern coastal boundary. We show that results from our modeling approach add validation to previous results and agree with the scaling relationship first introduced by \cite{Pollard-Rhines-Thompson-1973:deepening} and re-derived specifically for the upwelling context by HM. Finally, we present a discussion of our full upwelling source depth distribution using the results from model simulations with a set of passive tracers forced at the boundary for every model depth level; we then examine how this distribution depends on the strength of the wind stress and stratification (Section~\ref{sec:results}\ref{sec:results:distribution}).

\subsection{Simulating an upwelling zone with a freely-evolving source depth}
\label{sec:results:novel-simulations-aspects}

\begin{figure*}[tbh]
     \centering
     \begin{subfigure}[b]{0.65\textwidth}
         \centering
         \includegraphics[width=\textwidth]{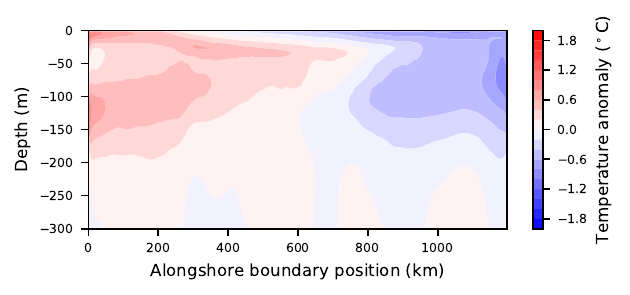}
         \caption{}
         \label{fig:theta_boundary}
     \end{subfigure}
     \hspace{0.01\textwidth}
     \begin{subfigure}[b]{0.3\textwidth}
         \centering
         \includegraphics[width=\textwidth]{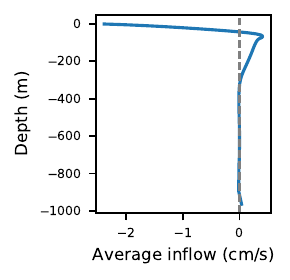}
         \caption{}
         \label{fig:inflow_boundary}
     \end{subfigure}
    \caption{(a) Temporal-mean temperature anomaly from the prescribed vertical profile along the western boundary of the physical domain just outside of the sponge layer. Temperature gradient with warmer temperatures to the south and cooler temperatures to the north is allowed to develop in this model configuration. (b) Vertical profile of the along-boundary averaged inflow. Vertical dashed line plotted at the zero velocity level. Strong offshore flow occurs at the surface and an onshore return flow is able to develop organically in the interior between 100 and 300 m depth. }
    \label{fig:boundary}
\end{figure*}

Unlike many previous studies with periodic boundary conditions in the alongshore direction, which require prescribing a body force to generate a balanced return flow, our results show that the modeling approach introduced here allows this return flow to develop organically. By using a model configuration with a sponge layer along the western boundary and restoring of the mean alongshore temperature in each model depth level, instead of using periodic boundary conditions and the commonly used point-by-point temperature restoring, the model is able to develop alongshore temperature and pressure gradients. Fig.~\ref{fig:theta_boundary} shows the time-averaged temperature anomaly from the prescribed vertical profile along the western boundary of the physical domain, at the inner edge of the sponge layer, for a simulation with medium wind stress and stratification. The model domain develops a clear meridional temperature gradient with warmer temperatures to the south and cooler temperatures to the north across a range of depths. These meridional gradients support the alongshore pressure gradient that geostrophically balances the sustained onshore return flow in the interior.  The along-boundary average of the inflow velocity in Fig.~\ref{fig:inflow_boundary} shows that, for a simulation with medium wind stress and stratification, the return flow primarily occurs just below the Ekman layer and extends to a depth of about 300 m. Critically, for the purpose of this work, the model determines the depth, magnitude, and time variability of the return flow on its own.

\begin{figure*}[tbh]
    \centering
    \includegraphics[width=\textwidth]{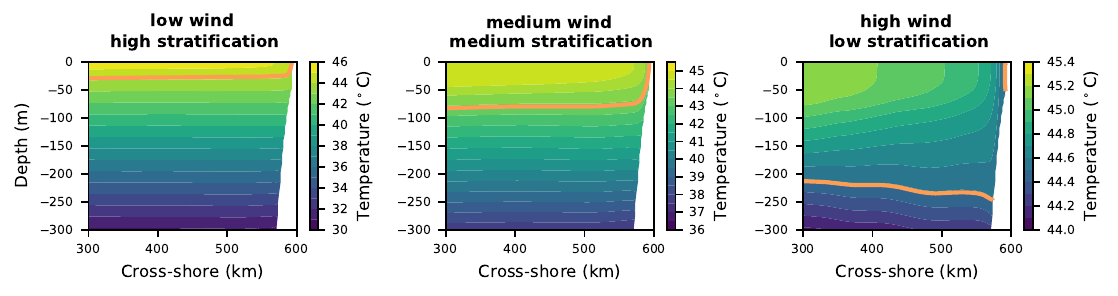}
    \caption{Cross-sections of the temporal-mean, alongshore-averaged temperature in three simulations. The contour interval and temperature range displayed depend on the stratification. A single orange contour highlights the isopycnal outcropping in the defined upwelling zone. Isopycnal outcropping occurs from greater depth for higher wind stress and lower stratification.}
    \label{fig:theta_cross}
\end{figure*}

The depth of the return flow should ultimately play a role in determining both the upwelling source depth and, correspondingly, the depth from which isopycnals outcrop. Cross-sections of temperature for three wind and stratification cases are shown in Fig.~\ref{fig:theta_cross} and illustrate that isopycnals outcrop from different depths depending on the strength of the wind stress and stratification \cite[as discussed by][]{Jacox-Edwards-2011:effects, He-Mahadevan-2021:how}. Stronger winds cause greater isopycnal outcropping for the same stratification. In all cases, far offshore, isopycnals flatten, and the alongshore mean temperature remains close to the prescribed vertical temperature profile. These results indicate that the model successfully allows the upwelling source depth, illustrated here by outcropping isopycnals, to vary freely while still maintaining the mean prescribed stratification.

\subsection{The mean source depth}
\label{sec:results:mean-depth}

\begin{figure*}[tbh]
    \centering
    \includegraphics[width=\textwidth]{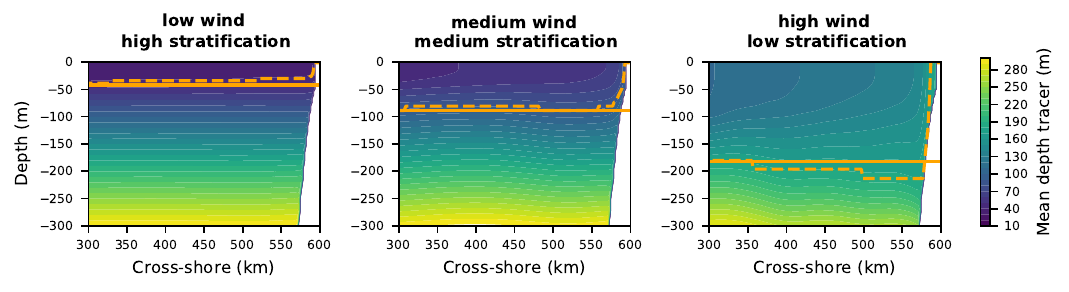}
    \caption{Cross-sections of the temporal-mean, alongshore-averaged mean-depth tracer in three simulations. The contour interval is 10 m. The solid orange line is drawn at the mean upwelling source depth given by the mean-depth tracer in the upwelling zone for each experiment. The mean upwelling source depth is greater (deeper) for higher wind stress and lower stratification. The dashed orange line highlights the contour of the mean-depth tracer that outcrops in the defined upwelling zone.}
    \label{fig:tracer_cross}
\end{figure*}

\begin{figure*}[tbh]
    \centering
    \includegraphics[width=\textwidth]{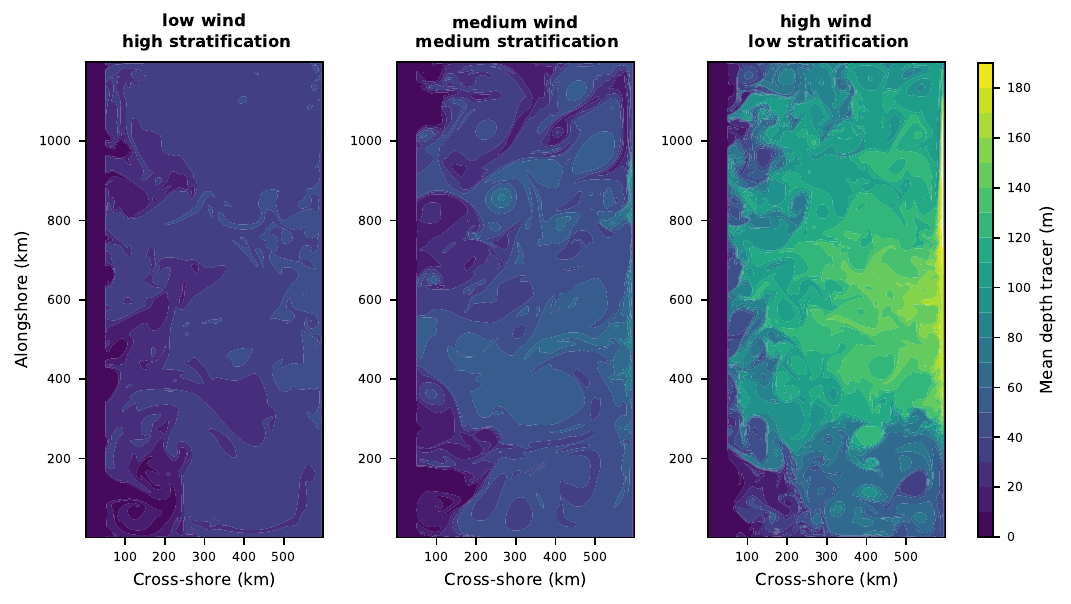}
    \caption{Snapshots of the mean-depth tracer at the surface in three simulations. The contour interval is 10 m. The surface waters with the deepest source depth occur in the upwelling zone.}
    \label{fig:tracer_surf}
\end{figure*}

The time-averaged results of our mean source depth tracer are shown by the vertical sections in Fig.~\ref{fig:tracer_cross}, and snapshots at the surface are shown in Fig.~\ref{fig:tracer_surf}. This tracer (see methods, Section~\ref{sec:methods}\ref{sec:methods:mean}) is restored along the boundary to the mean depth  in each vertical level. The value of this tracer in the upwelling zone, therefore, describes the mean depth from which the upwelling fluid originates and captures the mean contributions from different source depths along the domain boundary due to both advection and mixing. A given value of this tracer at the surface within the upwelling zone could mean, for example, that the water came from that depth exclusively, or that it is the result of waters from shallower and deeper depths mixing along the way to the surface. The upwelling source depth quantified by the value of the mean-depth tracer observed in the surface upwelling zone in Figs.~\ref{fig:tracer_cross} and \ref{fig:tracer_surf} ranges from 41 m to 182 m in the different experiments. The deep ocean water is then transported offshore by the surface Ekman transport, resulting in deep source waters distributed across large portions of the domain at the surface. Far offshore, near the surface as well as below the source depth, the mean source depth of fluid parcels in every depth level is nearly equal to their current depth, showing that, as expected, the flow outside of the upwelling zone is largely horizontal. We reiterate that, throughout the domain, the mean source depth of any given fluid parcel is affected by both advection and mixing, the separate effects of which are not apparent by studying the mean source depth tracer alone. In the next section (\ref{sec:results}\ref{sec:results:distribution}), we will address the contributions due to mixing by investigating the full source depth distribution.

We estimate the mean upwelling source depth as the time-mean value of the mean-depth tracer in the upwelling zone (defined in the methods, Section~\ref{sec:methods}\ref{sec:methods:mean}). It is also possible to estimate the mean upwelling source depth using the concentration of each of the 50 depth level tracers in the upwelling zone according to Equation~\eqref{eq:Ds_mean}. This calculation is similar to HM, except that their tracers were initialized at each level rather than forced at the boundary and therefore could not be used to examine the statistical steady state of the mean source depth. We find these two methods yield identical results with the obvious computational efficiency advantage of the single mean-depth tracer introduced here (Supplemental Fig.~1).

The upwelling source depth in our experiments ranges from 41 m in the low wind, high stratification case to 182 m in the high wind, low stratification case. Consistent with prior work \citep{Jacox-Edwards-2011:effects, He-Mahadevan-2021:how}, we find that, for a given stratification, the source depth increases (deepens) with increasing strength of the wind stress and, for a given wind stress, the source depth decreases (shallows) with increasing strength of the stratification (Fig.~\ref{fig:tracer_cross}). We note that a greater upwelling source depth does not necessarily correspond to denser (colder) upwelled waters; while the initial and boundary-restored surface temperature and density are the same in all simulations by construction, the prescribed stratification varies, and two experiments with the same mean source depth but different stratification strengths would result in surface waters of different densities in the upwelling zone. The density of the upwelled water increases with both increasing wind stress and stratification. We find that the greatest upwelling source depth occurs for high wind and low stratification (Fig.~\ref{fig:tracer_cross}) while the greatest density of upwelled waters occurs for high wind and high stratification \cite[Fig.~\ref{fig:theta_cross};][]{Jacox-Edwards-2011:effects, He-Mahadevan-2021:how}.

\begin{figure}[tbh]
    \centering
    \includegraphics[width=.4\textwidth]{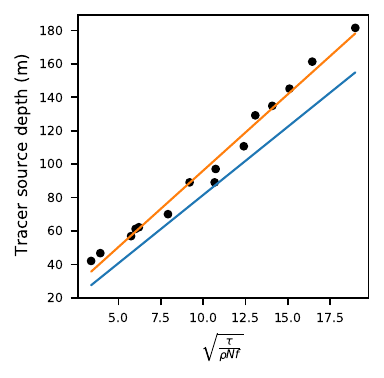}
    \caption{Mean-depth tracer estimate compared to Equation~\ref{eq:Ds_scaling} scaling for mean upwelling source depth. Blue line shows the theoretical estimate with intercept 0, $C_s$=8.16 in Equation~\ref{eq:Ds_scaling}. Orange line shows the line of best fit with intercept=4.83, slope=9.13 ($r^2=0.99$).}
    \label{fig:mean_scaling}
\end{figure}

Previous work suggests that the mean upwelling source depth, $D_s$, depends on the wind stress, stratification, density, and Coriolis frequency as described by the scaling relation,
\begin{equation} \label{eq:Ds_scaling}
    D_s = C_s\sqrt{\frac{\tau}{\rho_0Nf}}.
\end{equation}
HM derived this scaling relationship for the coastal upwelling source depth by assuming that the wind-driven circulation is balanced by the eddy-driven restratification in the coastal upwelling zone \citep{marshall2003residual, Thomsen-Capet-Echevin-2021:competition}. The scaling in Equation~\eqref{eq:Ds_scaling} was shown to hold in an idealized numerical upwelling model using periodic boundary conditions, flat bottom topography, and no surface heat flux \citep{He-Mahadevan-2021:how}. We find that, despite several non-trivial differences between modeling configurations, the results from our numerical model experiments are consistent with the scaling given by Equation~\eqref{eq:Ds_scaling} (Fig.~\ref{fig:mean_scaling}). HM further theorized that
$C_s=8.16$ by utilizing a previously estimated coefficient describing the strength of the eddy-driven streamfunction. While \cite{Pollard-Rhines-Thompson-1973:deepening} derived the same scaling for the depth of the wind-driven mixed layer independently, they found the proportionality constant to be much smaller in this context ($C_s=1.7$).

Fitting a line to this scaling with our model results for the mean upwelling source depth where both the slope and intercept are allowed to vary, yields a similar value of 9.13 for the slope ($C_s$) and an intercept of 4.83 with $r^2=0.99$ indicating a very strong fit. We suggest that there may not be justification for requiring the intercept of the relationship described by Equation~\eqref{eq:Ds_scaling} to be zero such that the relation is best written as $D_s = A+ C\sqrt{{\tau}/{\rho_0Nf}}$. In the case where the right-hand side of Equation~\eqref{eq:Ds_scaling} is zero (i.e., $\tau=0$), wind-driven upwelling is not expected to occur and, thus, the upwelling source depth is ill-defined. Additionally, the small difference in the slope of the relationship from HM may be affected by different modeling choices made in this study as compared to their work. In particular, the neglect of a surface heat flux in the numerical model used by HM may result in a smaller scaling slope for this relationship due to the following considerations. The assumption that the eddy-driven restratification compensates for the wind-driven circulation was shown to be valid only when there is no surface heat flux \citep{marshall2003residual, Thomsen-Capet-Echevin-2021:competition}. In a study of the competition between baroclinic instability and Ekman transport in the Southern Ocean, the presence of a surface heat flux was shown to decrease the strength of the eddy driven streamfunction \citep{Thomsen-Capet-Echevin-2021:competition}. A weaker eddy-driven streamfunction, and thus restratification, in the coastal upwelling context would imply a steeper slope for the relationship described by Equation~\eqref{eq:Ds_scaling}. In addition, bottom topography and shelf slope have previously been shown to affect the mean source depth \citep{Jacox-Edwards-2011:effects} and may also contribute to the relatively small differences in scaling slope observed here.

\subsection{The source depth distribution}
\label{sec:results:distribution}

\begin{figure*}[!tb]
    \centering
    \includegraphics[width=\textwidth]{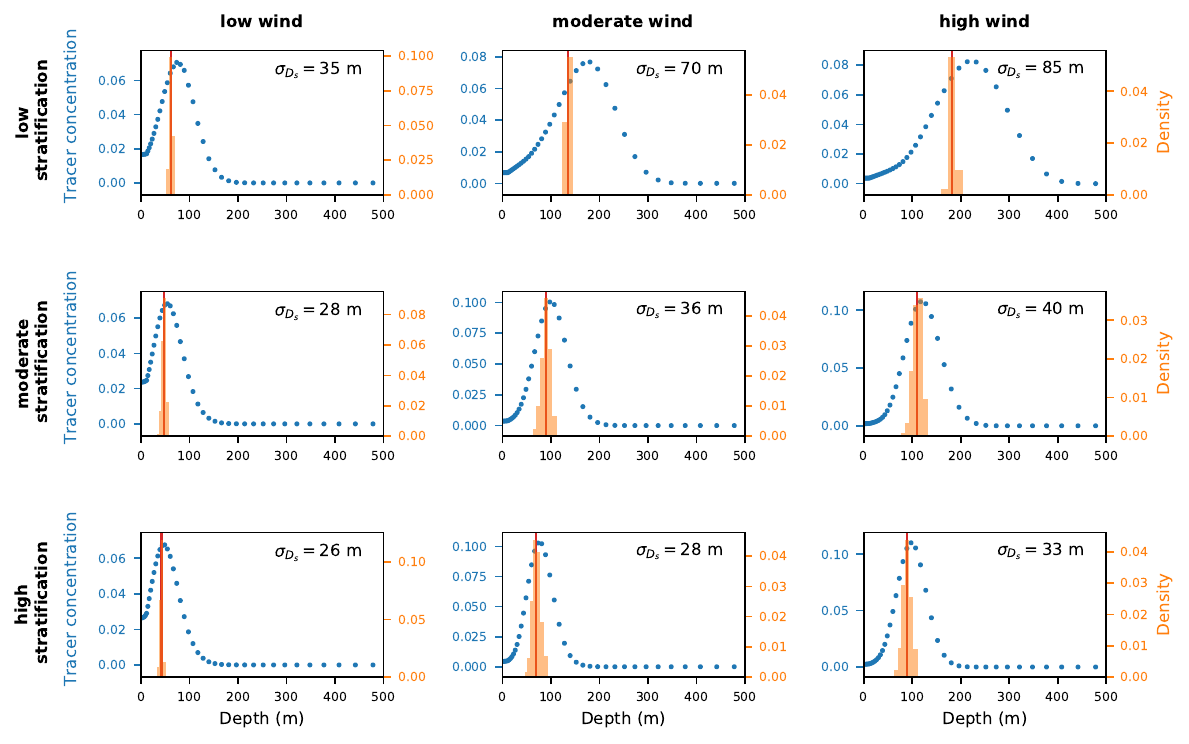}
    \caption{Comparison of the full upwelling source depth distribution informed by the 50 passive tracers to the distribution of the mean source depth over time in each of the nine main simulations. Temporal-mean, alongshore-averaged tracer concentration (density) plotted for each of the 50 model depth level tracers; blue points indicate the concentration of each corresponding depth tracer occurring in the surface upwelling zone. Transparent orange bars represent the the time-varying distribution of the alongshore-average of mean source depth. Solid vertical red line indicates the temporal mean, alongshore-averaged upwelling source depth.}
    \label{fig:depth_dist}
\end{figure*}

The mean upwelling source depth discussed above and in previous work only captures a single average source, while we actually expect the upwelled waters to originate from a range of depths due to various cross-isopycnal mixing processes. We therefore further advance the discussion of the source of upwelling by considering the full distribution of depths from which the water in the upwelling zone originates. We construct a full source depth distribution and quantify the center and spread of the distribution with the mean and standard deviation, respectively using our set of 50 passive depth tracers (Methods Section \ref{sec:methods}\ref{sec:methods:distribution}). We compute the temporal-mean, alongshore-average concentration of each depth level tracer in the previously defined upwelling zone, quantifying the contribution of multiple source depths to the upwelled waters at the surface and constructing the upwelling source depth distribution. This distribution, shown in Fig.~\ref{fig:depth_dist}, characterizes the extent to which the upwelled waters originate from a range of source depths.

Notably, there is substantial spread in the upwelling source depth distribution as shown by the widths of the distributions in the panels of Fig.~\ref{fig:depth_dist}. In particular, we find that, similar to the trends in the mean upwelling source depth, the width of the source depth distribution increases with increasing wind stress for a given stratification, and the width of the source depth distribution decreases with increasing stratification for a given wind stress. We propose the following explanations for these trends. Greater wind stress increases the strength of the upwelling and leads to steeper isopycnals and, therefore, to stronger eddy motions, which, in turn, lead to cross-isopycnal mixing that enhances the width of the depth distribution. Higher stratification leads to a narrow range of source depths since the increased stratification tends to suppress vertical motion and mixing. 

To quantify the width of the source depth distribution, which represents the range of depths that the upwelled waters originate from, we compute the standard deviation of the distribution using Equation~\eqref{eq:Ds_std}. Our modeling approach also allows us to calculate the time variability and the temporal standard deviation of the mean source depth (Fig.~\ref{fig:depth_dist}). We compare the source depth distribution informed by the 50 passive tracers to the distribution of the mean source depth over time and find that the time variance of the mean source depth is much smaller than the variance of the source depth distribution (Fig.~\ref{fig:depth_dist}). This suggests that the variance in the upwelling source depth distribution must be created through cross-isopycnal mixing. This can be seen by considering the case of time variability in the upwelling source depth due to internal variability in the upwelling zone, but without any cross-isopycnal mixing. Because there is no mixing in this scenario, each isopycnal level would correspond to exactly one initial source depth, and all of the variability in the depth tracers in the upwelling zone would be due to the time variability of the outcropping isopycnals. Thus, with no diapycnal mixing, the source depth distribution given by the time variability of the upwelling source depth would match that given by the depth tracers. The gap between the time variance of the mean source depth and the variance of the source depth distribution observed in all experiments as shown by Fig.~\ref{fig:depth_dist} must, therefore, be generated by cross-isopycnal mixing.

Cross-isopycnal mixing may occur due to several different physical processes: vertical diffusion in the interior, submesoscale mixing across near-vertical isopycnals in the upwelling region \citep{Capet-Marchesiello-McWilliams-2004:upwelling}, the breakup of filaments created by eddies and the 3-D mixing effects due to submesoscale subduction processes \citep{gula2022submesoscale}, or air-sea fluxes at the surface. Additional variance in the source depth distribution may also be created at the western boundary by injecting the passive depth tracers into multiple isopycnal levels. We further discuss the individual contributions to the spread of the source depth distribution of each of these mechanisms below.

\begin{figure*}[!tb]
    \centering
    \includegraphics[width=\textwidth]{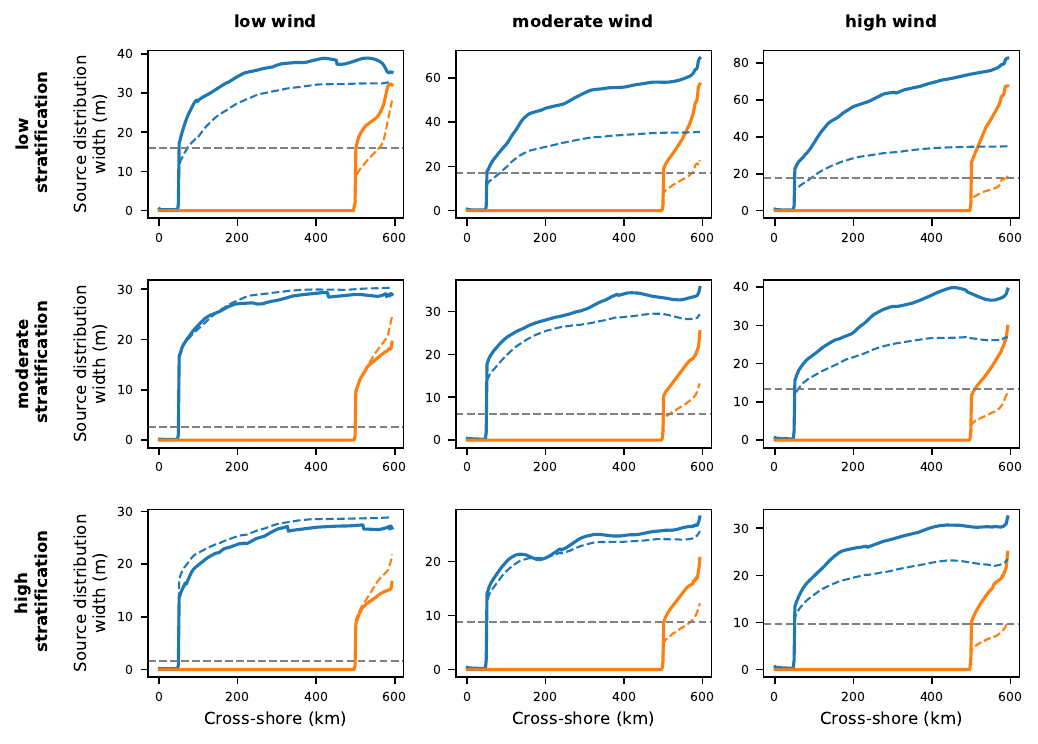}
    \caption{Standard deviation of the upwelling source depth distribution plotted along an idealized path of an upwelled parcel. (Path follows the mean upwelling source depth contour from the western boundary toward the coast and then up the continental slope at the eastern boundary into the upwelling zone. See dashed orange lines in Fig.~\ref{fig:tracer_cross}.) The solid blue line shows results from the control model run with the source region defined as the western boundary sponge layer. The solid orange line shows results from a model run with an added restoring region 100 km offshore where the depth and age tracers are strongly restored to their initial value. The horizontal dashed grey line represents the standard deviation in the distribution created by injecting tracers into multiple isopycnal levels in the western boundary sponge layer. The dashed lines represent the contribution to the standard deviation due to vertical mixing by diffusion alone matched on age and depth for experiments with the source region at the western boundary in blue and experiments with the source region 100 km offshore in orange. The contribution due to diffusion was estimated using 1-D simulations of diffusion according to Equation~\eqref{eq:sim_diffusion} and matched based on depth and age of the water parcel along the idealized path.}
    \label{fig:std_path}
\end{figure*}

To better understand the physical mechanisms responsible for the generation of the variance in the source depth distribution, we aim to quantify how the variance in the source depth distribution grows as a water parcel moves from the source region toward the upwelling zone. To do so, we construct an idealized path in the $x-z$ plain, representing the path that a water parcel would take if it followed a contour of the $y$-averaged mean-depth tracer all the way from the source region in the western boundary sponge layer to the upwelling zone. Specifically, to construct this path, we find the model depth level corresponding to the grid cell in which the mean-depth tracer is closest to the mean upwelling source depth at each cross-shore position, yielding a single depth at which the idealized water parcel travels for each cross-shore position. The idealized path can be seen in Fig.~\ref{fig:tracer_cross} where it is represented by the orange dashed line. We then compute the standard deviation of the source depth distribution at all grid points along this idealized path in each experiment (solid blue lines in Fig.~\ref{fig:std_path}). We find that the standard deviation of the source depth distribution along this path is typically the largest in the upwelling zone at the surface (at the end of the idealized path). This large variance is consistent with the expectation that the submesoscale eddies driving cross-isopycnal mixing in the outcropping zone mix water from different source depths that were carried along isopycnal surfaces toward the surface, ultimately leading to a larger variance there. But surprisingly, the standard deviation of the source depth is high and generally close in magnitude to that in the upwelling zone along most of the defined path, all the way to the western boundary (Fig.~\ref{fig:std_path}). The variance throughout the domain, away from the outcropping zone, may be generated away from the upwelling zone due to contributions from injecting the passive tracers into multiple isopycnal levels in the western boundary and by vertical diffusion in the interior. Alternatively the variance may be generated in the upwelling zone by submesoscale mixing across near-vertical isopycnals in the upwelling region and then transported horizontally along isopycnals across the domain by mesoscale mixing leading to the high variance observed throughout the domain. We investigate the magnitude of each of these contributions to the width of the source depth distribution below.

\begin{figure}[!tbhp]
    \centering
    \includegraphics[width=.4\textwidth]{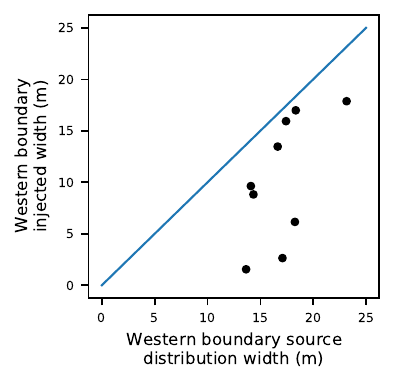}
    \caption{Comparison of the standard deviation of the source depth distribution along the western boundary estimated using passive tracers against the standard deviation of the source depth distribution due to injecting tracers at multiple isopycnal levels along the western boundary. The standard deviation from injecting tracers in the western boundary sponge layer was estimated from the variance of the temperature along the edge of the western boundary sponge layer using Equation~\eqref{eq:T_var}. Blue line is the one-to-one line. All data points fall below the one-to-one line; the standard deviation due to injecting tracers into multiple isopycnal levels is always less than the observed standard deviation of the source depth distribution at the western boundary.}
    \label{fig:T_std_wb}
\end{figure}

Due to the way the tracers are injected in the source region, some of the variance observed in the source depth distribution throughout the domain is generated at the western boundary. While each tracer is injected into a single model depth level, the depth of a given isopycnal may experience small magnitude variations due to the eddy variability in the isopycnal depth along the inflow boundary. This variability results in each isopycnal being injected with multiple different depth tracers over time, which will create variance in the source depth distribution at the western boundary. We note that while this may seem like an effect of the fixed western boundary source region in this model, the same effect would occur anytime a source region is explicitly defined, and a source region must be defined for the source depth to be calculated. We can estimate the variance contribution from injecting tracers at the western boundary to the source depth distribution by calculating the variance of the temperature in a given model depth level at the western boundary. The standard deviation of the source depth distribution is related to the temporal standard deviation of the western boundary temperature at the source depth as follows:
\begin{equation} \label{eq:T_var}
    \sigma_{D_s}(wb) = \frac{dT}{dz} \sigma_T(wb) = \frac{N^2}{g\alpha} \sigma_T(wb),
\end{equation}
where $\sigma_{D_s}(wb)$ is the standard deviation of the source depth distribution at the western boundary, $\sigma_T(wb)$ the temporal standard deviation of the temperature on the western boundary, and $\alpha=2\times10^{-4}$ K$^{-1}$ is the thermal expansion coefficient. The magnitude of the contribution to the variance of the source depth distribution due to injecting tracers at multiple isopycnal levels is represented by the dashed grey line in Fig.~\ref{fig:std_path}. We directly compare the standard deviation of the source depth distribution at the mean upwelling source depth just outside the sponge layer near the western boundary to that generated by injecting the tracers at the boundary in Fig.~\ref{fig:T_std_wb}. The variance generated by injecting tracers along the western boundary at the source depth is consistently less than the total variance there, showing that this mechanism alone does not explain the large variance in the source depth distribution in the upwelling zone and throughout the domain.

\begin{figure*}[tbh]
    \centering
    \includegraphics[width=\textwidth]{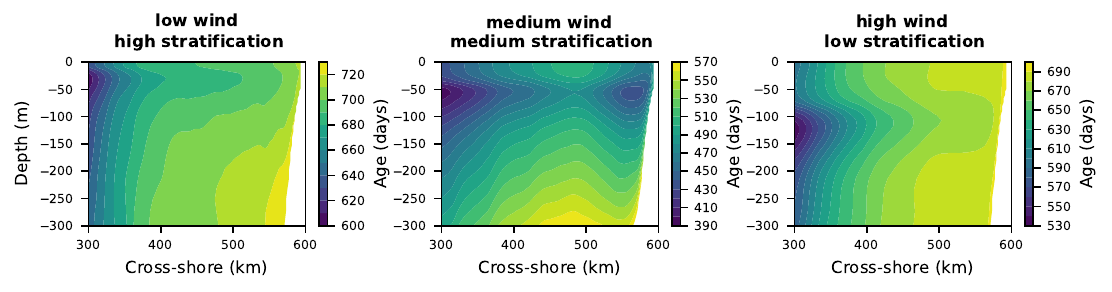}
    \caption{Cross-sections of the temporal-mean, alongshore-averaged age tracer in three simulations. The source region (where the age is 0) is defined to be the western boudnary sponge layer. The contour interval and age range displayed depend on the experiment.}
    \label{fig:age_cross}
\end{figure*}

Next, we estimate the contribution of vertical mixing to the variance of the source depth distribution by performing 1-D simulations of a diffusion equation applied to each of the 50 passive depth tracers separately. For each tracer, we initialize the concentration of the tracer to 1 in its corresponding depth level and 0 elsewhere (identically to how it is initialized and forced within the western boundary in the full model, see Methods Section \ref{sec:methods}\ref{sec:methods:distribution}). Then we simulate the vertical profile of each passive depth tracer due to one-dimensional vertical diffusion according to
\begin{equation} \label{eq:sim_diffusion}
    \frac{\partial C_k}{\partial t} = \kappa \frac{\partial^2 C_k}{\partial z^2}.
\end{equation}
Here, $C_k$ is the concentration of the passive tracer for depth level $k$ and $\kappa=10^{-5}$ m$^2$/s is the same value for the eddy diffusion coefficient used in the full numerical model. We use the same discretization of the vertical levels in this 1-D simulation as in the full numerical model and impose no flux boundary conditions at the surface and bottom of the domain. The results of these simulations emulate the evolution of the expected concentration of each of the 50 passive tracers in each vertical level over time due to diffusion. Using these results, we can thus estimate the standard deviation in the source depth distribution (Equation~\ref{eq:Ds_std}) due to diffusion at any depth level as a function of travel time from the source region. We estimate the travel time of each water parcel in the nine main simulations using a passive age tracer that is restored to zero in the source region in the western boundary sponge layer (Fig.~\ref{fig:age_cross}) and thus measures the travel time of fluid parcels from this western boundary source region. The travel time from the source region tells us the total time a given parcel was affected by diffusion and, therefore, we can evaluate the standard deviation of the source depth distribution due to diffusion alone from the results of the 1-D diffusion simulations. Overall, we find that diffusion explains a non-negligible fraction of the variance in the source depth distribution and, in fact, is responsible for almost all of the variance generated in the low wind simulations (dashed lines in Fig.~\ref{fig:std_path}). However, in cases where the upwelled water is drawn from greater depths (stronger winds and weaker stratification), diffusion explains less than half of the total variance. These results suggest that there must be another source of variance generation in the source depth distribution and, in particular, more variance is created in the source depth distribution when the upwelling source depth is greater.

To further resolve where the variance in the source depth is generated, we perform the following mechanism denial experiment. We add an additional alongshore restoring layer beginning 100 km offshore and extending to the western boundary where, in each model depth level, the concentration of the corresponding depth-level tracer is restored to one, and all other tracer concentrations including the age tracer are restored to zero, akin to the restoring in the western boundary sponge layer. The values of the depth and age tracers therefore are now related to a source region located in this sponge layer 100 km offshore rather than at the western boundary. In these simulations (orange lines in Fig.~\ref{fig:std_path}), we find that the standard deviation of the source depth distribution is substantially higher near the upwelling zone compared to at the edge of the source region 100 km offshore. This shows that there is consistently a substantial contribution to the standard deviation of the source depth distribution from the mixing processes occurring in the upwelling zone. Specifically, strong submesoscale mixing across outcropping isopycnals in the upwelling zone explains the variance in the upwelling source depth distribution generated near the coast. Together with the results of the 1-D vertical diffusion simulations, these results suggest that while the most substantial contribution to variance in the source depth distribution is generated in the upwelling zone, other mechanisms for generating variance, such as vertical diffusion, cannot be ignored.

The importance of vertical mixing due to diffusion in the above results also suggests that, ultimately, the variance of the source depth distribution in the upwelling zone depends on where the source waters are defined. For example, we note the larger standard deviation of the source depth distribution for the simulations run with a single restoring layer at the western boundary compared to those run with an additional restoring layer 100 km offshore from the coast. This gap can be explained by differences in the time available for vertical mixing to occur due to the water parcels taking a longer time to arrive in the upwelling zone from the source region when the restoring layer is farther from the coast. The farther from the upwelling zone that the source region is defined, the longer diffusion has time to act, and the more dominant the contribution of vertical mixing by diffusion to the generation of source variance will appear. However, the overall patterns in which the standard deviation of the source depth distribution increases with increasing wind stress and decreases with increasing stratification remain consistent regardless of where the source waters are defined.

\section{Discussion} \label{sec:conclusion}

We have presented a discussion of the upwelling source depth and, importantly, the full upwelling source depth distribution. While previous work \citep{Jacox-Edwards-2011:effects, He-Mahadevan-2021:how} has focused primarily on the mean source depth, we expect that the source waters actually originate from a range of depths, and the full distribution of sources may have important implications for setting the SST and determining the nutrient content of the upwelled water. To flexibly model an eastern boundary coastal upwelling system in a statistical steady state, we developed an idealized numerical modeling approach that, unlike those used in some previous efforts, does not prescribe a body force at a set depth to generate a geostrophycally balanced return flow and allows this return flow to evolve organically. This means that the model can determine the depth of the return flow and the source depth of the upwelling, which also makes it possible to examine how the upwelling source depth distribution evolves. We then computed the mean upwelling source depth using a proposed single passive depth tracer and constructed the source depth distribution using a unique passive tracer for every model depth level that tracks the contribution of each model depth level to the upwelled waters. These tracers are forced within the source region at the western boundary of our regional ocean model.

We have shown that our numerical modeling approach provides an idealized representation of the coastal upwelling dynamics that is consistent with the main observed features of eastern boundary upwelling systems. We found that a previously developed scaling relationship for the mean upwelling source depth described by Equation~\eqref{eq:Ds_scaling} \citep{Pollard-Rhines-Thompson-1973:deepening, He-Mahadevan-2021:how} holds despite different assumptions and modeling approaches. Having constructed the full upwelling source depth distribution, we quantified the width of the distribution using the standard deviation and found similar trends in the width of the source depth distribution as have been previously established for the mean source depth--increasing source depth distribution width with increasing wind stress and decreasing width with increasing stratification. We discuss how the variance in the source depth distribution is created by several processes including by injecting tracers into multiple isopycnal levels along the western boundary, by cross-isopycnal mixing forced by vertical diffusion throughout the interior, and especially by submesoscale eddies mixing across near-vertical isopycnals within the upwelling zone near the surface. We found that, while cross-isopycnal mixing by vertical diffusion away from the upwelling zone contributes to the variance of the source depth distribution, it cannot explain all of the variance observed and, therefore, we concluded that a significant part of the variance in the upwelling source depth distribution must be generated due to cross-isopycnal submesoscale mixing in the upwelling zone.

We note that there are other factors that may play a role in determining the source depth distribution that we have not explored in this work. First, we force our model with a temporally and spatially uniform wind stress. Curl-driven upwelling may also occur over a broader region offshore when spatial gradients are present in the wind stress field \citep{Song-Miller-Cornuelle-et-al-2011:changes}. The temporal variability of the wind stress and specifically the occurrence of strong wind events driving coastal upwelling have been shown to impact the strength of coastal upwelling and may also play a role in setting the source depth \citep{botsford2006effects, garcia2010observations, Li-Luo-Arnold-et-al-2019:reductions}. Additionally, we only test one bottom topography profile, prescribe a single formulation for the surface heat flux, and impose a linear stratification via our boundary forcing, all of which are idealized and may affect the source of the upwelled waters in the coastal upwelling system. Finally, we defined the source region as the western boundary sponge layer of the model but there could still be inflow into the middle third of the domain from the north and south. This approach allows us to simply estimate the source depth distribution from an offshore source, but in a realistic system, one might be interested in source regions to the north and south of the upwelling region, which we cannot examine using our model setup.  We leave further study of more realistic choices to future work.

Nonetheless, our investigation of the upwelling source depth distribution contributes to the understanding of the source of the upwelled waters in the coastal upwelling system beyond the mean source depth. Our results have important implications for the resulting sea surface temperatures and the upper ocean nutrient content relating to their dependence on the strength of the wind stress and stratification. Using the framework developed here, one can quantify the mean and standard deviation of the source depth distribution in realistic simulations of upwelling systems to better understand and predict the state of coastal upwelling systems in past and future climates. For example, our results suggest that during the Pliocene warm period (approximately 3--5 million years ago), when the upwelling favorable wind stress is believed to have been weaker \citep{wara2005permanent, arnold2016reductions}, we would expect that both the mean upwelling source depth would have been shallower and the width of the source depth distribution would have been narrower implying warmer sea surface temperatures and a narrower nutrient distribution. This may have interesting implications to the proxy record derived from biological proxies that depend on the local nutrient content and may therefore depend on the source depth distribution width. Similarly, in our currently warming climate, where both the upwelling favorable wind stress \citep{bakun1990global, snyder2003future} and stratification \citep{mcgowan2003biological} are expected to increase, the HM scaling for the mean source depth described by Equation~\eqref{eq:Ds_scaling} and trends observed for the spread of the upwelling source depth distribution can help us to predict and understand how the upwelling source waters will change.


\clearpage
\acknowledgments

This publication uses results from the CASE project, operated by Scripps Institution of Oceanography and funded by the Climate Observations and Monitoring Program, National Oceanic and Atmospheric Administration, U.S. Department of Commerce. We would like to acknowledge high-performance computing support from Cheyenne (doi:10.5065/D6RX99HX) provided by NCAR's Computational and Information Systems Laboratory, sponsored by the National Science Foundation. This work was funded by NSF grant 2303486 from the P4CLIMATE program. ET thanks the Weizmann Institute for its hospitality during parts of this work. 


%
%
\datastatement
The dataset on which this paper is based is too large to be retained or publicly archived with available resources. The numerical model simulations upon which this study is based can be shared upon request to the corresponding author.

%






%



\clearpage
\newpage
\bibliographystyle{ametsocV6}
\bibliography{references}

\end{document}